\documentclass[sigconf, nonacm]{acmart}
\usepackage{float}
\usepackage{tabularx}
\usepackage{multirow}
\usepackage{balance} 
\usepackage{subfigure}

\AtBeginDocument{%
  \providecommand\BibTeX{{%
    \normalfont B\kern-0.5em{\scshape i\kern-0.25em b}\kern-0.8em\TeX}}}

\setcopyright{none}
\renewcommand\footnotetextcopyrightpermission[1]{}
\begin{document}

\title{You Have a Point There: Object Selection Inside an Automobile Using Gaze, Head Pose and Finger Pointing}


\author{Abdul Rafey Aftab$^{1,2,3}$, Michael von der Beeck$^2$, Michael Feld$^3$ }
\affiliation{$^1$ University of Saarland, Saarbr\"ucken, Germany }
\affiliation{$^2$ BMW Group, Munich, Germany}
\affiliation{$^3$ German Research Center for Artificial Intelligence (DFKI), Saarbr\"ucken, Germany }
\affiliation{\texttt{\{abdul-rafey.aftab, michael.beeck\}@bmw.de, michael.feld@dfki.de}}

\renewcommand{\shortauthors}{Abdul Rafey Aftab, Michael von der Beeck, Michael Feld}

\begin{abstract}
Sophisticated user interaction in the automotive industry is a fast emerging topic. Mid-air gestures and speech already have numerous applications for driver-car interaction. Additionally, multimodal approaches are being developed to leverage the use of multiple sensors for added advantages. In this paper, we propose a fast and practical multimodal fusion method based on machine learning for the selection of various control modules in an automotive vehicle. The modalities taken into account are gaze, head pose and finger pointing gesture. Speech is used only as a trigger for fusion. Single modality has previously been used numerous times for recognition of the user's pointing direction. We, however, demonstrate how multiple inputs can be fused together to enhance the recognition performance. Furthermore, we compare different deep neural network architectures against  conventional Machine Learning methods, namely Support Vector Regression and Random Forests, and show the enhancements in the pointing direction accuracy using deep learning. The results suggest a great potential for the use of multimodal inputs that can be applied to more use cases in the vehicle. 
\end{abstract}



\maketitle


\section{Introduction}

With the evolution of the technology for user interaction, the focus of the interface has gradually shifted from being computer-centered to being human-centered. In the last few decades, we have seen a great interest in the use of eye-tracking for Human-Computer Interaction (HCI) \cite{quek1995eyes, zhai1999manual, jacob1990you}. The use of eyes as an input method to track the users' gaze direction provides a natural interface without having touch-based inputs. However, the robustness of the eye-tracking methods depend on various factors such as the sensors used (e.g. head-mounted tracking device or a non-contact tracking device). Eye-gaze for user interaction imposes some restrictions as it can be very volatile and has an `always-on' characteristic. Therefore, it lacks a natural trigger for object selection. 

\begin{figure}[t]
  \includegraphics[width=0.48\textwidth]{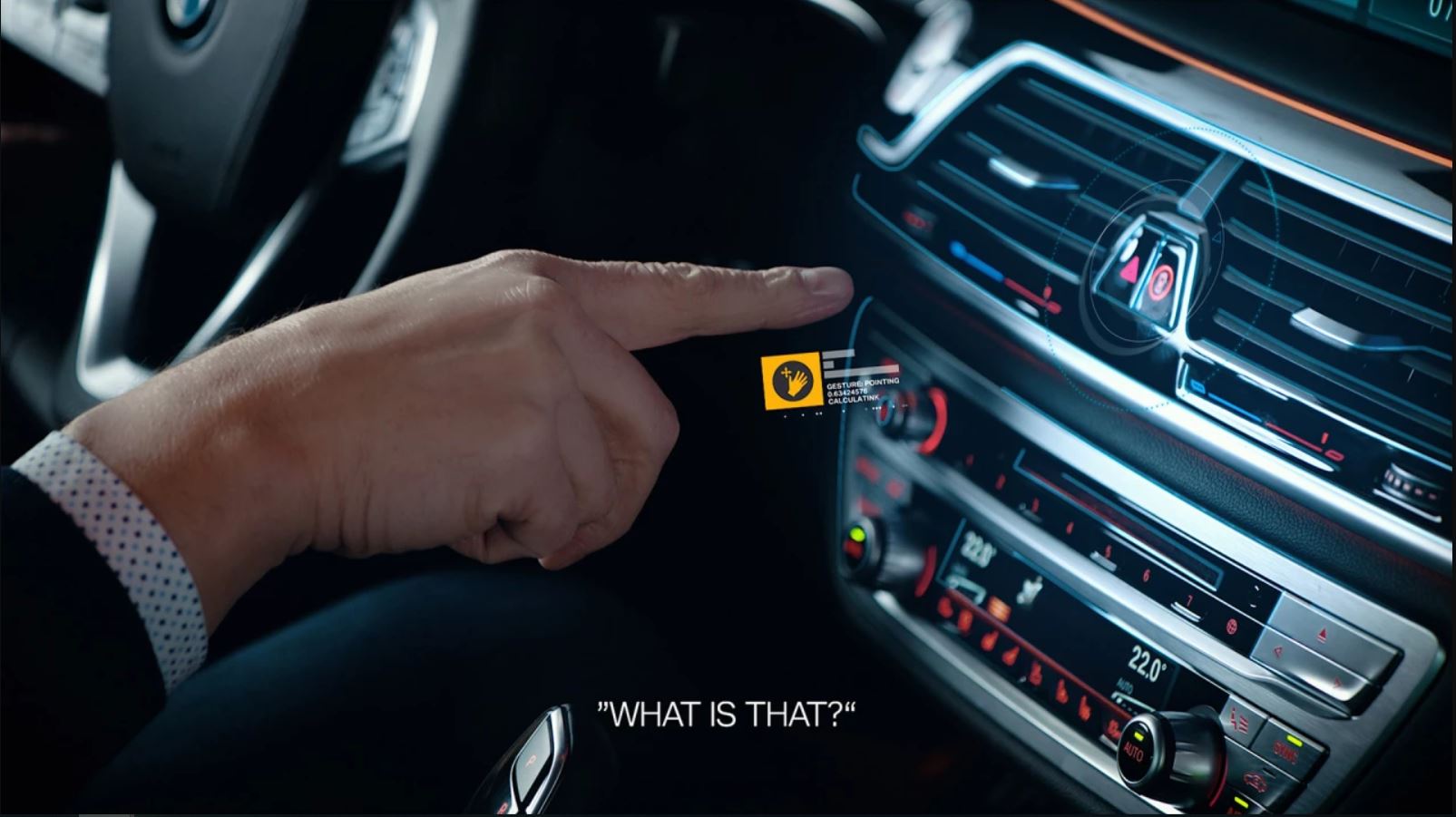}
  \caption{Driver makes a pointing gesture to interact with the car. Image courtesy of \cite{bmwnatint}}
  \label{fig:teaser}
\end{figure}

Another important and natural source of input for user interaction are mid-air gestures. Since early interactive systems, such as Bolt's seminal work ("Put that there" \cite{bolt1980put}), mid-air  pointing gestures have also been of significant interest for user interaction \cite{walter2014cuenesics, schweigert2019eyepointing}. This is because they enable users to point to and reference objects that are too far away to touch in a natural manner, especially with the help of speech commands \cite{sauras2017voge}. 

In this paper, we integrate the deictic information from gaze and head pose, along with a specific gesture, i.e., finger pointing gesture, in order to combine the efficiency and naturalness of these modes of interaction, using a late fusion approach. There are two main reasons for this multimodal integration: 1) to compensate the drawbacks in one modality using the other and 2) to enhance the overall performance of the user interaction by taking advantage of the temporal relations between them.
Although the use of multiple inputs for interaction can have numerous applications such as for assistance of people with disabilities, we focus our work on the use of head pose, eye-tracking and finger pointing for applications in the automobile industry. The motivation is driven by the direct application in the BMW Natural Interaction, presented at the Mobile World Congress 2019, that enables genuine multimodal operation for the first time \cite{MWC}.

While driving, the driver's cognitive attention is affected by numerous tasks such as navigating the infotainment system, operating the in-vehicle control units etc. It has been observed that the gestural interfaces are fairly easy to use and they reduce the cognitive load on the driver \cite{riener2012gestural, rumelin2013free}. For this reason, intuitive interactions via free hand pointing gestures are becoming increasingly prevalent for in-vehicle infotainment systems, which is particularly useful for novice users \cite{ahmad2017does}. Furthermore, multimodal interfaces also have a tendency to reduce the driver load while driving \cite{manawadu2017multimodal}. We, therefore, propose an in-vehicle user interaction, that supplements the touch-based inputs, in which the driver is able to operate control modules in the vehicle in a touchless and natural manner that is also comparatively less distracting. In order to identify the desired object or Area-of-Interest (AOI), the user may use a finger pointing gesture, as this type of gesture provides a deictic reference to the various real-world objects, as shown in Figure \ref{fig:teaser}. The action to be performed on the selected object may  be provided by speech commands, such as, "what is \textit{that}?" or "close \textit{that} window". 

In the context of this paper, we focus on the recognition of the object that the user selects with a combination of gaze, head pose and finger pointing. It has been observed that drivers make relatively large errors in pointing \cite{brand2016pointing, roider2018implementation}, and the integration of gaze improves the accuracy of pointing while driving \cite{roider2018see}. This is because, while driving, the eye-gaze is mainly focused on the road, but momentarily moves and fixates on the target object. In the past, it was observed that gaze anchoring to a target existed for the entire duration of the pointing movement of the finger \cite{neggers2001gaze}. More recently, Ahmed \textit{et al.} show that in most cases, the driver first looks towards the desired object before making the pointing gesture and that there exists a misalignment between the gaze and finger movements \cite{ahmad2017does}. We exploit this relation between the eye-gaze and the finger pointing motion to accurately choose the object located in the entire field-of-view of the driver. Furthermore, we include head pose as well as it is directly linked with eye-gaze \cite{jha2016analyzing}. 

\section{Related work}
Multimodal user interaction has a wide variety of applications for in-vehicle functions.
Mitrevska \textit{et al.} demonstrate the use cases of an adaptive multimodal control of in-vehicle functions with the help of an individual modality (speech, gaze or gesture) or a combination of two or more modalities \cite{mitrevska2015siam}.

Apart from the applications, research has shown that the use of multiple input modalities have a potential to outperform systems with a single input modality  \cite{liu2018efficient, esteban2005review, turk2014multimodal}. Due to this, the fusion of multiple modalities has been used by many researchers for user interaction. A multimodal technique for selection of objects on the screen, namely the MAGIC pointing, was presented by Zhai \textit{et al.} about two decades ago \cite{zhai1999manual}. This technique allows the user to select objects on a screen by fixating on the target and pressing a regular manual input device to trigger the selection. The Midas-touch problem faced in this technique can be overcome by using mid-air gestures to trigger the selection as presented in \cite{schweigert2019eyepointing, chatterjee2015gaze+, nesselrath2016combining}. Gaze is often used for selection of objects on a screen while using an additional trigger such as using a speech command \cite{maglio2000gaze}. However, selection performed exclusively by gaze is difficult, especially when objects are placed very close to each other \cite{hild2019suggesting}.

EyePointing is an extension to MAGIC pointing that uses finger pointing as a trigger for object selection \cite{schweigert2019eyepointing}. Chatterjee \textit{et al.} demonstrated a better outcome with the integration of gaze and gesture as inputs as compared to systems with only gaze or gesture  \cite{chatterjee2015gaze+}. Similarly, Nesselrath \textit{et al.} use a combination of three modalities, speech, gaze and gestures, to initially select objects of the vehicle, e.g., side mirrors or windows, and then use gestures or speech to control these objects \cite{nesselrath2016combining}. 

These approaches primary use the gaze information and enhance the naturalness of the user interaction with secondary modality, e.g. gestures or speech. In contrast to this, Sauras-Perez \textit{et al.} propose the use of speech with the finger pointing gesture for selection of Points-of-Interest (POI) while driving a vehicle \cite{sauras2017voge}. 

Unlike these approaches, we do not use finger pointing gesture as a trigger for selection, but rather use the finger ray-cast in addition to gaze ray-cast and head pose to fuse them together.
The fusion of the modalities is used for better performance while using speech as the trigger. Each modality is treated equally at the input, and the model learns the weights on its own during the training process.

The problem of object selection inside a car has also been presented by Roider \textit{et al.} who integrate eye gaze with finger pointing gestures in a passive manner using a simple rule-based fusion approach. They have shown that the selection on an in-vehicle display screen achieves increased pointing accuracy over the single modality, i.e., finger pointing \cite{roider2018see}. This experiment is limited to only four objects on a screen adjacent to each other. In our work, we enhance the object selection using a late fusion approach to select a wide range of objects inside the vehicle that lie in the hemisphere in front of the driver. We use an additional modality, i.e. head pose, because the head pose and gaze direction are mostly directly related and are usually considered together for recognizing the visual behaviour \cite{ji2002real, mukherjee2015deep}. Another reason for including head pose is that eyes can be easily occluded when using eye-tracking sensors that are fixed for practical reasons rather than head-mounted. Head pose may prove to be beneficial in such cases.

Head pose and gaze direction are used by Jha and Busso, in their work, to estimate driver's gaze direction \cite{jha2016analyzing}. They use simple linear regression models to predict the direction of gaze. Compared to this, the novelty in our method is that we allow the neural network to learn the relations between the three modalities, and then use linear regression to predict the output.

A driver query system, similar to our experimental setup, is presented by Kang \textit{et al.} in which the smart car is able to figure out where the driver is looking at using visual cues, head pose and speech  \cite{kang2015you}. We show that the accuracy of the driver's query can be further improved tremendously by adding finger pointing. However, we do not consider speech as an input to the fusion as it is a very different problem altogether, but rather use speech as a trigger for fusion. The dialogue duration has an impact on the interaction between the car and user \cite{strayer2014measuring}, which we do not explore.

Deep neural networks have been applied to feature fusion for various tasks \cite{ngiam2011multimodal}, but they usually deal with abstract data or abstract features. In \cite{olabiyi2017driver}, Olabiyi uses Deep Recurrent Neural Networks to perform a fusion of sensory inputs to predict driver actions. We use a similar concept and apply Convolutional Neural Networks (CNN) on the sequential (temporal) input data.

To sum up, we extend and combine concepts from various previous research for a more robust performance that can be practically applied in a real car. While many studies have been performed in simulators, only a few are presented in a real-world scenario. For practical reasons, our experiments are performed in a real car to provide the users a more realistic impression, and, therefore, to achieve more significant results than a simulation.

\begin{figure}[t]
    \centering
        \includegraphics[trim=0 120 0 0,clip, width=\linewidth]{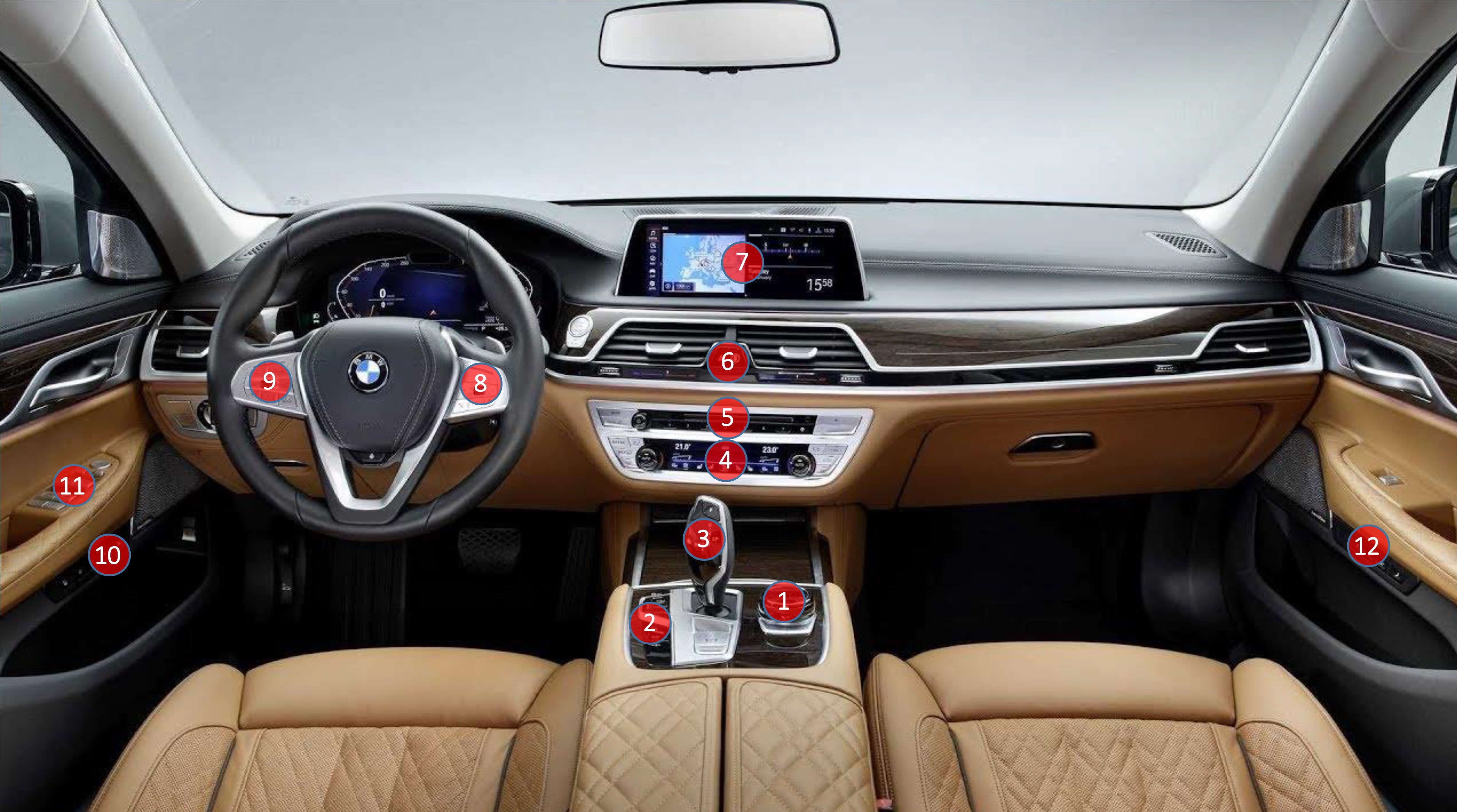}
        \caption{The 12 AOIs in the cockpit }
            \label{fig:AOI_car}
\end{figure}

\section{Data collection}
We consider a simple yet a very productive cockpit use case, that is object selection inside the vehicle. We chose to set up our sensors for data collection in a stationary but functional BMW vehicle rather than a simulator to have apt uses in real world. This is why we do not use head-mounted sensors as they may hinder the driver's cognitive abilities while driving.

We do not consider the action to be performed in the context of this work.  The selection of objects outside the car is excluded in this work. It is a future extension of this work, and we hope to integrate outside use cases in the future as well, e.g., pointing to buildings or landmarks, and inquiring about them.

\subsection{Apparatus}
 There are two types of camera systems used to capture the 3D information of the driver:  the Gesture Camera System and the Visual Camera System.
 
 \subsubsection{Gesture Camera System} 
The gesture camera, mounted next to the Roof Function Centre of the car, captures hand and finger movements in the 3D space using a Time-of-Flight (ToF) camera. It has a wide Field-of-View so that it covers almost the entire operating zone of the driver. The gesture camera system detects a finger pointing gesture and calculates the vector from the tip of the finger to the base of the finger. The 3D coordinates of the fingertip are used as the finger position.

\subsubsection{Visual Camera System}
The visual camera is a high-definition camera with a built-in technology that evaluates the images of the driver and calculates the required 3D vector data for the head pose and eye-tracking. It uses an actively illuminated infrared (IR) sensor to capture eye and head movements even in low lighting. The camera is placed behind the steering wheel in such a way that eye-occlusion is minimal, and it does not interfere with the driver's attention on the road. The algorithm integrated into the camera system calculates the head rotation as three Euler angles (roll, pitch and yaw), and the 3D coordinates of the estimated centre position of the head. In addition to this, it provides eye position which is the 3D coordinates of the cyclops eye (i.e., centre point between the two human eyes), and the 3D vector coordinates of the eye direction merged together from both the left and the right eye.

\begin{figure}[t]
    \centering
        \includegraphics[width=\linewidth]{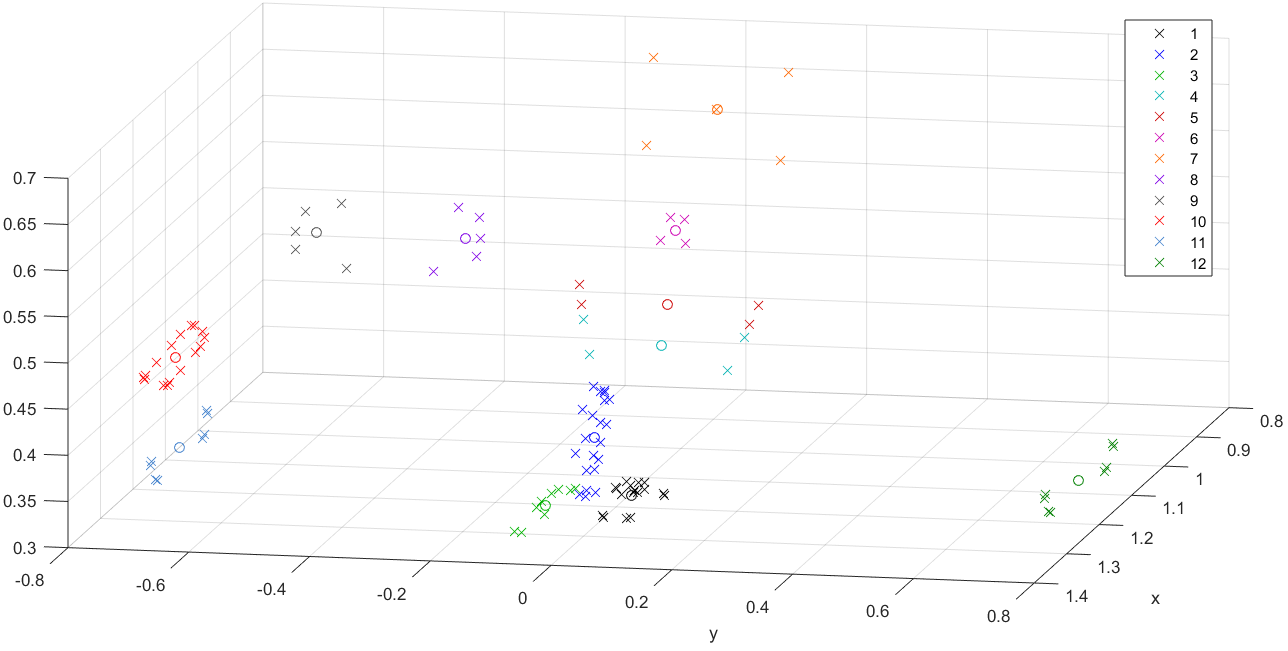}
            \caption{3D scatter plot of the measured AOI points. }
            \label{fig:scatter_AOI}
\end{figure}

\subsubsection{Speech command}
Apart from the visual camera and the gesture camera systems, we use a speech command along with the pointing gesture which we implemented with the Wizard-of-Oz (WoZ) method. In order to record the timestamp of the pointing gesture, a secondary person (acting as the wizard) pushes a button manually that stores the timestamp at the instant when the primary user points to an object and says, "what is \textit{that}?" The aim was to note the timestamp when the word "\textit{that}" is said. However, there may be human error involved in measuring this timestamp.

\subsubsection{Selection of AOIs}
Based on the use-cases for the driver interaction, we chose twelve distinct, but closely situated, control modules in the cockpit of the car as Areas-of-Interest (AOIs). They are illustrated with red circles in Figure \ref{fig:AOI_car}. A 3D scatter plot of all the measured points for each of the 12 AOIs is illustrated in Figure \ref{fig:scatter_AOI}. The `x' represent measured points, whereas the `o' shows the mean of the measured point of the AOI.

\subsection{Participants}
In this experiment, we collected data from 22 participants aging between 20 and 40 years old. 5 of these participants were female, and the remaining 17 were male. The drivers were asked to point to the various AOIs in a stationary vehicle and give the command, "what is \textit{that}?" There was no further instruction provided so that the pointing gesture can be as natural as possible. The participants were free to choose either hand and use any finger for pointing. 15 \% samples of pointing samples were performed with left hand while the remaining were   with the right hand. About 30 \% of participants wore glasses, 10 \% wore contact lenses, and the remaining had no glasses or lenses.

\subsection{Dataset Statistics}
From each of the 22 drivers, we collected 10 pointing gesture events for each of the 12 AOIs. The entire dataset consisted of 120 samples for every participant where each data sample consists of exactly one gesture pointing event. In total, we had 2640 samples that we collected for the cockpit use case. Due to errors in recording, 60 samples were discarded. Therefore, 2580 samples were used.

The difference between the estimated and actual direction, calculated for each of the modalities, is termed as the  estimation error. An error in estimation is considered when the vector direction is outside the visible surface area of the AOIs. The mean and standard deviation of the estimation error for the three modalities in the horizontal direction (azimuth) and the vertical direction (elevation) are shown in Table \ref{Table:AOIs}. It can be seen that the eye direction has relatively large errors for the first three AOIs, while the head direction has relatively large errors for AOIs 10 and 11. The large errors in the elevation angle of the head direction suggest that the head movement in the vertical direction is considerably small even when looking downwards (as almost all of the AOIs lie below the car windscreen). 

This is a relatively small dataset, especially when using deep neural networks. However, the outcomes that we achieved from such a small dataset (in section \ref{results}), show a great proficiency in our approach, which may even be enhanced using a larger dataset. Consequently, a much larger dataset that will be used in the future is being collected which incorporates more use cases as well.

\begin{table}[b]
\centering
\footnotesize
 \begin{tabularx}{\linewidth}{ c| c c | c  c |c c  }

 \multirow{3}{*}{AOI}
  & \multicolumn{2}{c|}{Eye Direction} & \multicolumn{2}{c|}{Head Direction} & \multicolumn{2}{c}{Finger Direction} \\
  \hline
 & \textbf{Azimuth} & \textbf{Elevation} & \textbf{Azimuth} & \textbf{Elevation} & \textbf{Azimuth} & \textbf{Elevation}  \\
  & \textbf{M (SD)} & \textbf{M (SD)} & \textbf{M (SD)} & \textbf{M (SD)}& \textbf{M (SD)} & \textbf{M (SD)}   \\  
 \hline
1 & 26°  (18°) & 13°  (11°) & 5°   (7°) & 37°  (10°) & 17° (35°) & 15° (17°) \\
2 & 23°  (17°) & 11°  (11°) & 4°   (6°) & 36°  (10°) & 11°  (22°) & 7°  (14°) \\
3 & 25°  (16°) & 19°  (13°) & 3°   (8°) & 46°  (11°) & 15° (30°)  & 17°  (17°) \\
4 & 2°   (5°)  & 1°    (4°) & 2°   (5°) & 23°   (8°) & 1°  (7°)   & 4°  (8°) \\
5 & 1°   (4°) & 1°   (5°)   & 1°   (4°) & 22°   (8°) & 1°  (5°)   & 5°  (8°) \\
6 & 5°   (5°) & 3°    (2°)  & 7    (8°) & 21°   (8°) & 9°     (16°) & 8°  (11°) \\
7 & 1°   (3°) & 1°    (2°)  & 3°   (7°) & 10°   (7°) & 1°      (6°) & 1°  (5°) \\
8 & 3°   (6°)  & 3°    (6°) & 5°   (5°) & 32°  (11°) & 3°  (10°)   & 4°  (8°) \\
9 & 2°   (4°)  & 3°    (6°) & 21° (12°) & 33°  (14°) & 9°   (19°) & 6°  (11°) \\
10 & 12° (13°) & 4°    (7°) & 34° (14°) & 36°  (14°) & 29°  (29°) & 26° (23°) \\
11 & 25° (14°) & 9°   (11°) & 45° (14°) & 31°  (13°) & 27°  (28°) & 25° (24°) \\
12 & 17° (18°) & 4°    (5°) & 11° (14°) & 15°   (7°) & 10°  (26°) & 6° (11°)

\end{tabularx}
\caption{Mean (M) and Standard Deviation (SD) of the estimation error (in degrees) of eye, head and finger direction. }
\label{Table:AOIs}
\end{table}

\begin{figure}[t]
   \begin{subfigure}
   \centering
    \includegraphics[trim=0 0 0 40,clip, width=\linewidth]{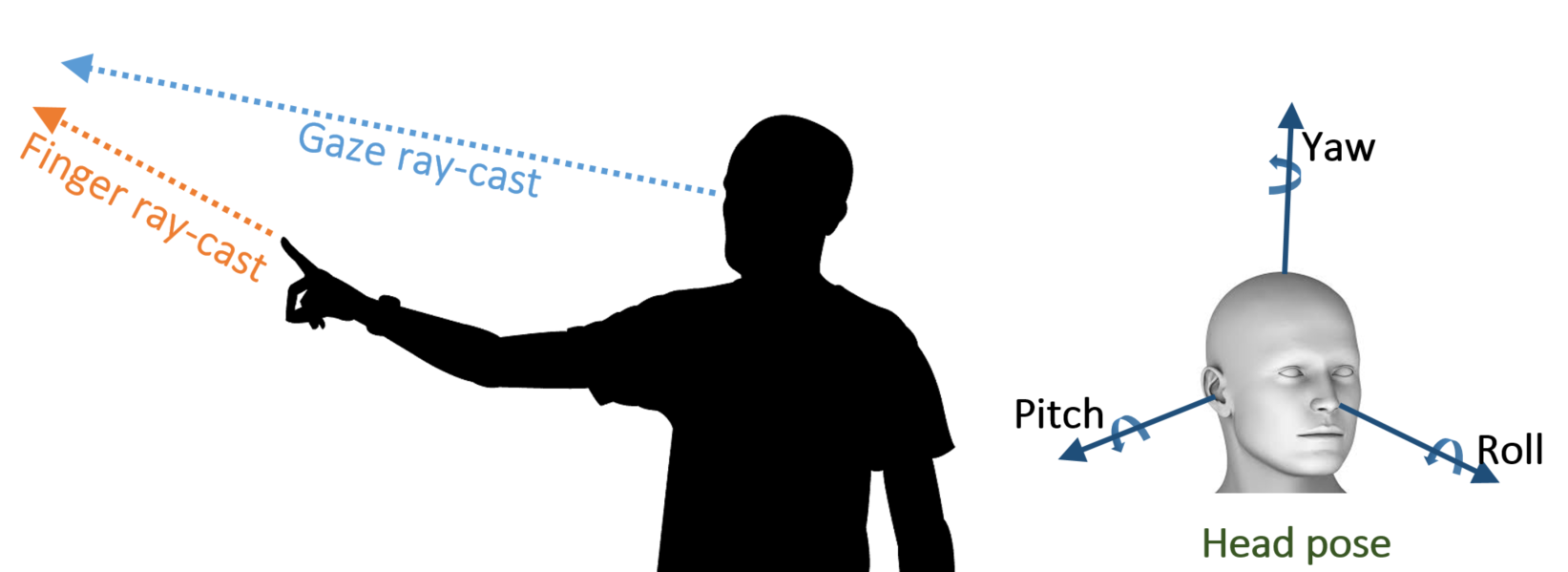}
    \caption{The modalities used: gaze and finger ray-cast (left) and head pose (right)}
    \label{fig:rays}
\end{subfigure}

\begin{subfigure}
\centering
    \includegraphics[trim=30 0 0 0,clip, width=\linewidth]{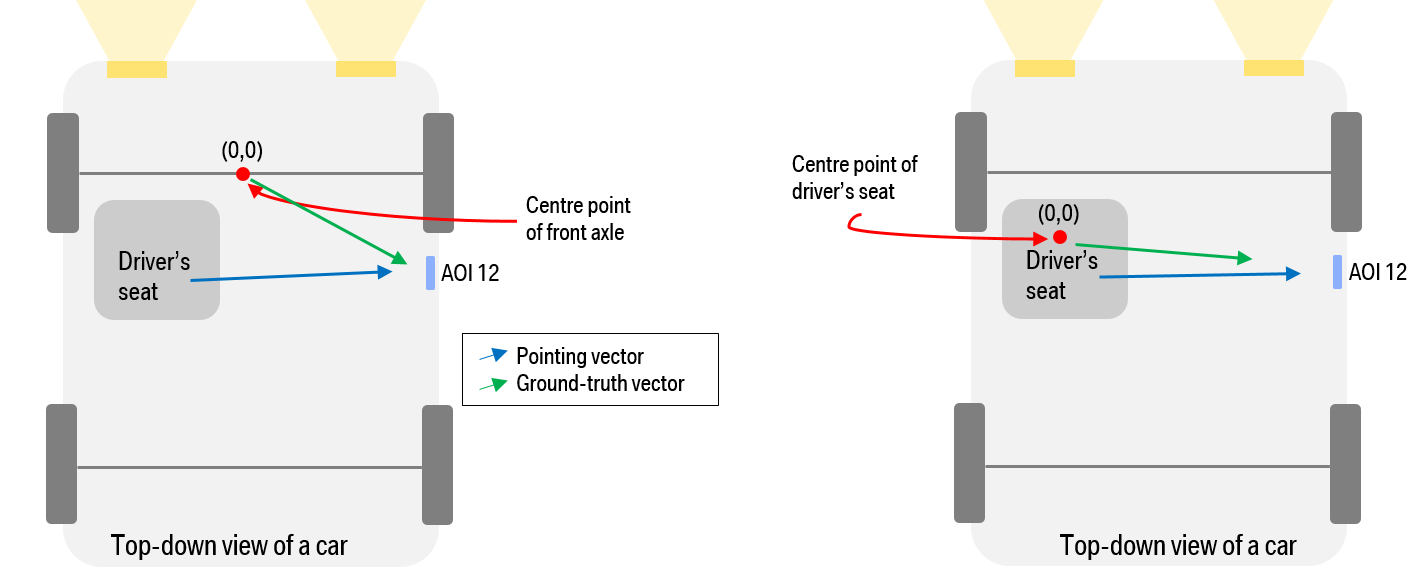}
    \caption{Change of origin from centre of front axle of the car (left) to the centre of the driver's seat (right)}
    \label{fig:origin}
\end{subfigure}
\end{figure}

\section{Methodology}
\begin{figure*}[t]
    \includegraphics[width=0.97\linewidth]{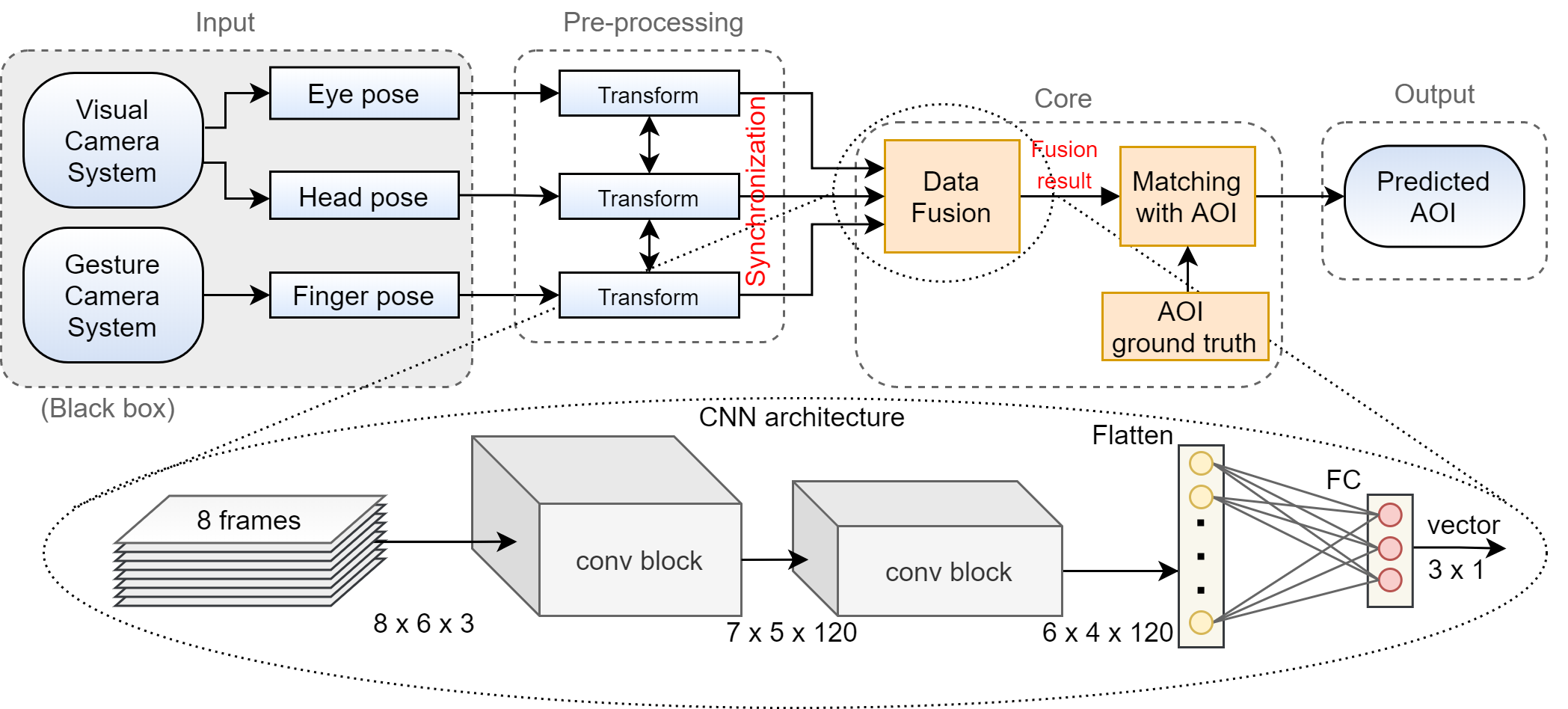}
    \caption{A multimodal late fusion architecture}
    \label{fig:methodology}
\end{figure*}

Our proposed architecture is given in Figure \ref{fig:methodology}. The camera systems are treated as black boxes that provide input processed from the images. A late fusion is applied on the preprocessed data to estimate the driver's referenced object by matching to the known AOIs. 

\subsection{Preprocessing}
The visual camera and the gesture camera systems provide the head, eye and gesture pose. Pose consists of both position and direction/rotation of the individual modalities. The finger and eye-gaze directions use ray casting while, for the head rotation, we use yaw, pitch and roll as inputs, as shown in Figure \ref{fig:rays}. The direction vectors of eye and finger are normalized to have a unit norm. Due to occlusion of the eyes or the finger, there are some frames with missing data. Occlusion of the eyes mainly occurs when the driver looks downward, and therefore, the eyelids occlude the pupils, or when the pointing arm comes in front of the face. To fill the missing data, we use linear interpolation from the two nearest neighbouring frames. Afterwards, camera calibrations are applied to give real-world coordinates with respect to the origin, i.e., the centre of the front axle of the car, which is the ISO standard \cite{sayers1996standard}. We apply a translation in order to translate the origin point to the centre of the driver's seat as illustrated in Figure \ref{fig:origin}. We observed from experiments that using such a translation makes the learning process in neural networks slightly better, i.e., the translation lead to an increased accuracy.

We evaluated different time intervals to figure out the right interval to use as input. The results are omitted in this paper. It was found out that choosing a small time period of 0.2 seconds at the instant when the WoZ button is pressed provided sufficiently good results. At about 45 fps, this time period amounts to 8 frames from each sample, 4 frames before the noted timestamp by the wizard, and 4 frames after it.

\subsection{Fusion Algorithm}
We use machine learning methods, particularly Deep Neural Networks (DNN), to fuse the three modalities. The motivation behind the choice is that there may be a number of different cases that may be difficult to address with a rule based approach. DNNs, with supervised learning, have a tendency to tackle the different cases on their own, provided that the dataset has a large variance. Additionally, DNNs are easily expandable to add more use cases which we will consider in the future work as well. We present a base model and show the comparison of results to other similar models. 

\subsubsection{Ground Truth Definition} 
The corner points of the AOIs in the car were measured, w.r.t to the origin (i.e. the centre of the driver's seat), as illustrated in Figure \ref{fig:scatter_AOI}.
We then define the ground truth as the 3D vector (with unit norm) calculated from the origin to the mean point of the measured points. The mean of the measurements reveal to be very close to the actual centre of the AOI (see Figure \ref{fig:scatter_AOI}). 

\subsubsection{Base model}
The input to the model, $x$, is a batch of samples of size $b$, such that $x \in \mathbb{R}^{b \times f \times a \times d}$, where $f$ is the number of frames used, $a$ is the number of feature attributes used and $d$ is the number of dimensions in each attribute. We chose 6 feature attributes as inputs: the position and direction of eye, head and finger. Each of these has 3 dimensions, representing a point or a vector in the 3D vector space. The inputs from 8 frames are concatenated together. The base model is a deep Convolutional Neural Network (CNN), consisting of two 2D convolutional layers and one fully connected layer. The output is linearly regressed to give a $(3 \times 1)$ vector for the fused direction. 

We use the cosine similarity  between the predicted vector and the ground truth vector as the loss function, i.e., the cosine of the angle between the two vectors. The loss function is, thus, given as:
\begin{equation}
    \mathcal{L} = \frac{1}{N} \sum^N_{i=1} \text{cos} (\theta_i) 
    = \frac{1}{N} \sum^N_{i=1} \frac{\textbf{\^y}_i \ . \ \textbf{y}_i}{\|\textbf{\^y}_i \| \  \| \textbf{y} _i\|} 
    \label{eqn}
\end{equation}

where $ \textbf{ \^y }_i$ is the $i$-th predicted fusion vector, $\textbf{y}_i$ is the $i$-th ground truth vector, $\theta_i$ is the angle between the two 3D vectors, and $N$ is the total number of samples. Therefore, we have $\mathcal{L} \in [-1, 1]$.

\subsubsection{Other models}
The base CNN model is compared with other models to show the performance of the fusion using various approaches. These include a  Fully-Connected Neural Network (FC-NN), Recurrent Neural Network (RNN), Support Vector Regression (SVR), and Random Forests (RF) regression.  The RNN has 2 LSTM (Long Short-Term Memory) layers and 1 fully connected layer. The FC-NN consists of 3 fully connected layers. Both, FC-NN and RNN, use the cosine similarity as the loss function. Conventional machine learning approaches, namely SVR and RF, are evaluated to figure out if the model complexity with DNNs is an overkill. For the SVR, the polynomial kernel is used with degree 2. 

\subsection{Matching Predictions with AOIs}
\label{matching}
In order to identify the desired object by the user, we measure the cosine similarity between the predicted vector and the 12 AOIs separately. The one with the highest cosine similarity is chosen. In other words, the one with the lowest angular deviation from the predicted value is chosen.

\begin{figure*}[t]
    \includegraphics[width=\linewidth]{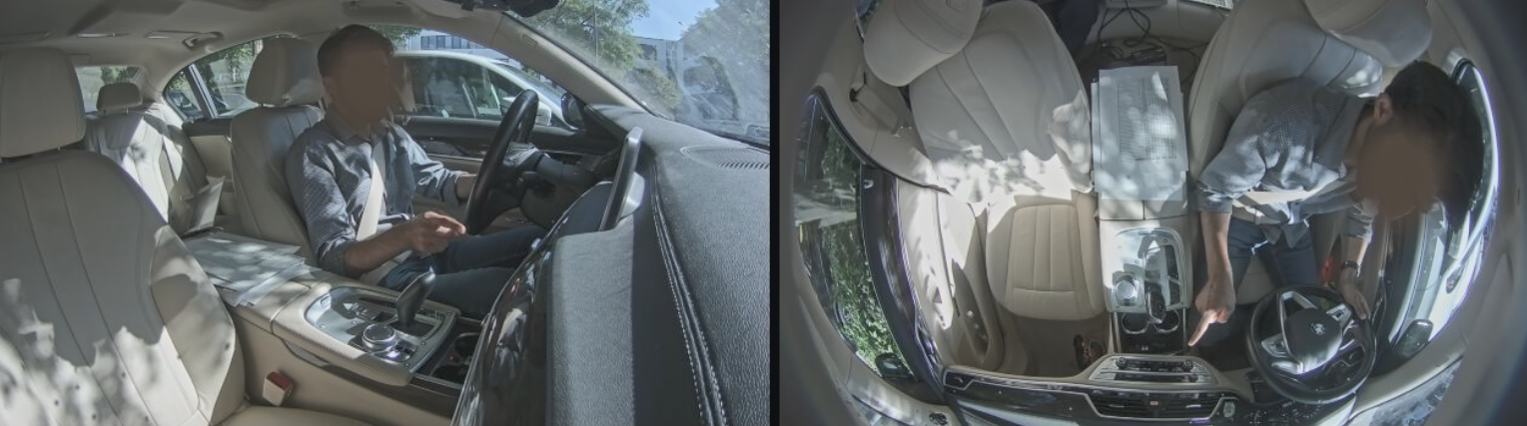}
    \caption{Driver points to AOI 5 (multimedia) in the car. The face is blurred due to the privacy policy.}
    \label{driver_pointing}
\end{figure*}

\section{Experiments and Results}
\label{results}
In the context of this paper, we report only one use-case, i.e., the cockpit use case. Figure \ref{driver_pointing} shows an example of the driver pointing to one of the AOIs.
Other experiments are in progress for further use cases, and the results will be presented in future publications. 

Due to the small size of the dataset, we use cross validation to report the results. A leave-one-out cross-validation  (22-fold cross-validation) is used for training and evaluation of the dataset. The dataset is split into three, training, validation and testing. With the leave-one-out cross validation, the test set covers the entire dataset.
The train/test split of the data is user-based, i.e., no sample from the participants in the training set appears in either the validation or the test set and vice versa. This means that each fold of the data contains 120 samples ($\approx$ 5\%) from each driver, respectively. 

We train the model for 100 epochs using a batch size of 8 and the Adam optimizer with a learning rate of 0.001. The best performing model on the validation set is selected and tested on the test set in order to avoid overfitting. The test results on the 22 folds are averaged to give the final value. By averaging using the cross-validation, we avoid the bias in the results which may occur due to certain user specific referencing.

\subsection{Metrics}
For the training and test loss, we use the cosine similarity function as shown in Equation \eqref{eqn}. For performance measures, we use two metrics: accuracy and Mean Angular Deviation (MAD).

\subsubsection{Accuracy} We use accuracy to evaluate the classification performance of the models for the 12 AOIs. Accuracy is the percentage of the correctly identified finger pointing samples:
\begin{align}
    \text{Accuracy} = \frac{TP}{{N}} \times 100 \%
\end{align}
where $N$ is the total number of predictions and $TP$ is the total number of true predictions (or correct identification) by the model.

\subsubsection{Mean Angular Deviation (MAD)} We define MAD as the mean of the angles between the predicted vectors and the corresponding ground truth vectors in the 3D vector space. This evaluates the performance of the regression output. Consequently, the lower the MAD, the better. Mathematically, we have:
\begin{align}
    \text{MAD} = \frac{1}{N} \sum^N_{i=1} \theta_i = \frac{1}{N} \sum^N_{i=1} \text{arccos}\left( \frac{\textbf{\^y}_i \ . \ \textbf{y}_i}{\|\textbf{\^y}_i \| \  \| \textbf{y} _i\|} \right)  
\end{align}

\subsection{Ablation Study}
In order to see the effect of each individual modality, we first do an ablation study. The ablation study also includes the removal of the difficult cases of AOIs, and analyzing the results. For all the experiments in the ablation study, the base model (CNN) is used with a 22-fold cross-validation.

\subsubsection{Effect of removing modalities}
\begin{figure}[t]
    \includegraphics[width=0.89\linewidth]{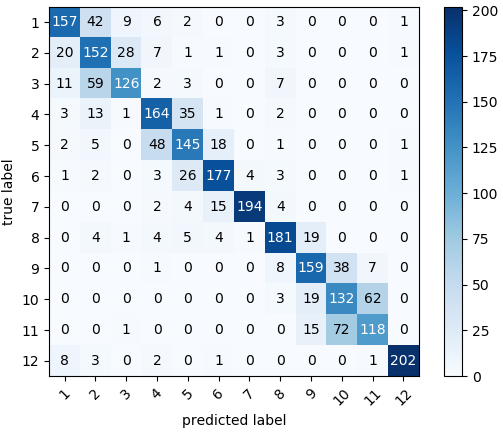}
    \caption{Confusion matrix for fusion of all three modalities for the 12 classes}
    \label{fig:conf_mat}
\end{figure}

The training consists of either one modality (e.g. position and direction of eye only), a combination of two modalities or a fusion of all three modalities. The cross-validation scores are shown in Table \ref{Table:monomodal}. It can be seen that the finger pointing accuracy (64.5\%) is significantly higher than the other two modalities. One possible reason for the low accuracy for the gaze (38\%) might be given by the missing gaze data. In the data, we found 662 samples out of 2640 to have the gaze data missing at the instant when the wizard pushes the trigger. Half of the missing gaze data occurs for the AOI 1, 2 and 3 for which the driver needs to look downwards to the right, while a quarter of these occur for AOI 10 and 11. The effect can also be seen in the confusion matrix shown in Figure \ref{fig:conf_mat} where AOIs 1, 2 and 3 as well as AOIs 10 and 11 have many misclassifications.

From the results, we see that the main contributor is the finger pointing gesture. By adding an additional modality on top of finger, there is a  slight increase of about 4\% in performance. Moreover, using all the three modalities, increases the accuracy of further by about 4 - 5\%. In a way, the head pose compensates for the missing eye gaze information and vice versa to improve the monomodal resultant of finger pointing.  

\subsubsection{Effect of removing difficult classes}
As it was observed that there was a significant amount of gaze data that was missing for the AOIs 1, 2, and 3, we removed these samples from our dataset to see the effect of gaze. There were 1940 samples that remained in total. The results are shown in Table \ref{Table:monomodal2}.

\begin{figure}[t]
    \includegraphics[width=0.89\linewidth]{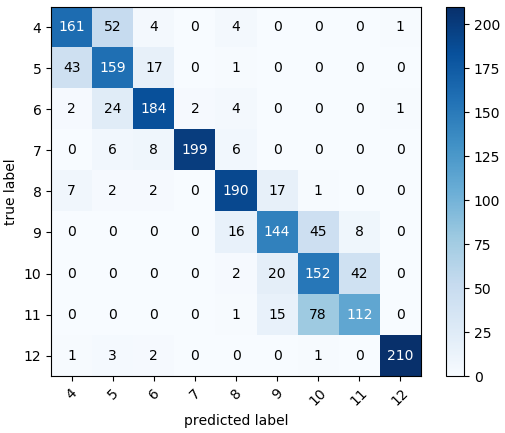}
    \caption{Confusion matrix for fusion of all three modalities for the 9 classes}
    \label{fig:conf_mat_9}
\end{figure}

It is observed that using these 9 AOIs (AOI 4 - 12) only, the eye-gaze accuracy significantly improves from 38\% to 57\%. The MAD also improves with a decrease of 1.6°. This demonstrates that the data collected for the AOIs 1, 2 and 3 has some errors. The accuracy of fusion (of all three modalities) also increases by about 4\%. The confusion matrix for the fusion is shown in Figure \ref{fig:conf_mat_9}. The most misclassifications are seen between AOI 10 and 11 which are very close together on the bottom-left side of the driver. A possible reason may be occlusion of the eyes from the arm when the driver uses the right hand for pointing. 

\begin{figure}[H]
    \includegraphics[width=0.883\linewidth]{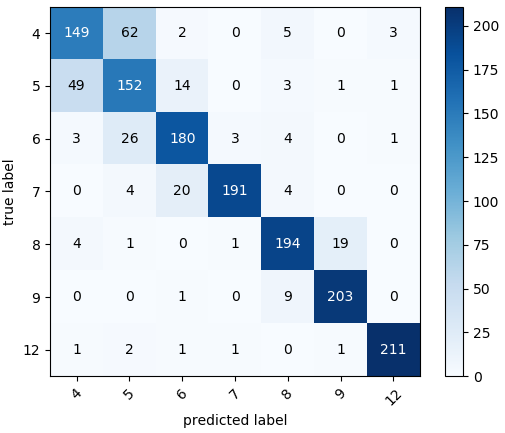}
    \caption{Confusion matrix for fusion of all three modalities for the 7 classes}
    \label{fig:conf_mat_7}
\end{figure}

The model accuracy and MAD after further removing AOIs 10 and 11 can be seen in Table \ref{Table:monomodal3}. The confusion matrix, for the model that uses all modalities, is shown in Figure \ref{fig:conf_mat_7}. The accuracy by using only gaze information as input, is increased by 8\% to 65\%, while the accuracy by using all modalities increases by 6\% to 83.9\%.

\subsection{Model Performance on Individual Drivers}
\begin{figure}[H]
    \includegraphics[width=0.95\linewidth]{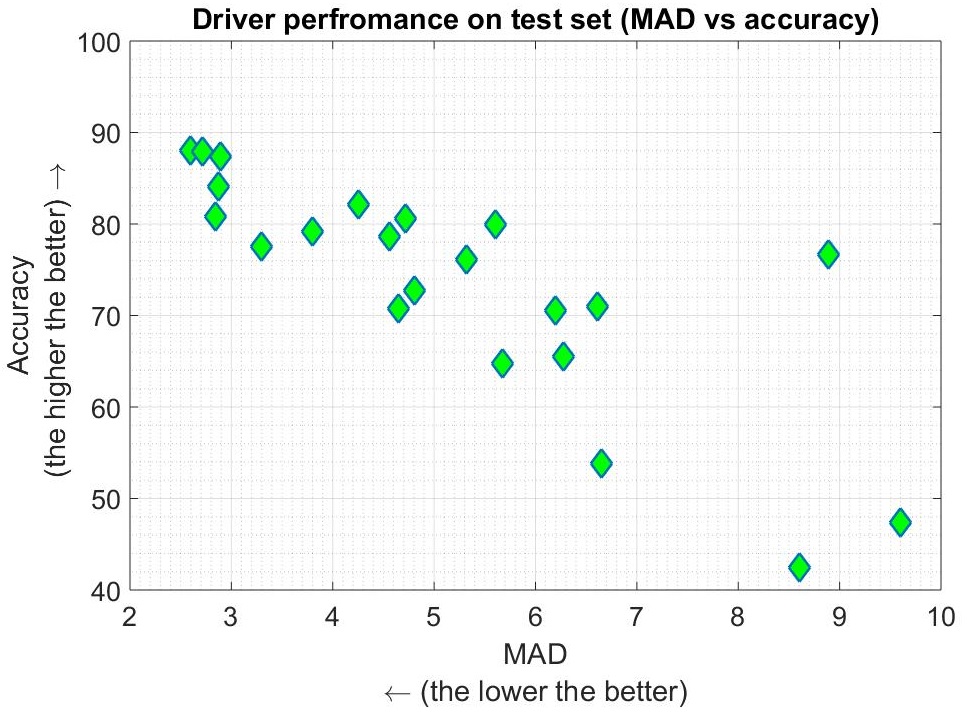}
    \caption{Model performance on the individual drivers}
    \label{fig:driver_test}
\end{figure}

We observed that different driver's point in a very different way. The scatter plot of performance of the CNN based model on the different drivers in the test set is illustrated in Figure \ref{fig:driver_test}. We plotted the accuracy on the `y' axis and MAD on the `x' axis. Therefore, the best point on the plot would be top-left and the worst point would be bottom-right.
We found out that accuracy and MAD of testing the base model on some participants is more than 80\% and less then 3°, respectively. On the other hand, test results on a few other participants demonstrate significantly poor pointing performance, i.e., less than 50\% accuracy and more than 8° MAD. This can be seen at the bottom left corner of the Figure \ref{fig:driver_test}.
Upon removing the two users from the dataset, and performing a 20-fold cross-validation, the model achieved a test accuracy of 75.1\% and a MAD of 5.4°. There is only a slight increase because the dataset becomes even smaller with 20 users.

\begin{table}[H]
 \begin{tabular}{|c | c c | }
 \hline 
 \textbf{Modality} & \textbf{Accuracy} $\uparrow$ & \textbf{MAD} $\downarrow$  \\  
 \hline\hline
 Head & 30.4 \%& 11.7°  \\ 
 \hline
 Gaze & 38.1 \%& 9.8° \\
 \hline
 Finger & 64.5 \%& 9.7° \\
 \hline\hline
 Gaze + Head & 50.5 \%& 7.7°  \\
 \hline
 Finger + Head & 68.5 \%& 6.5°  \\ 
 \hline
 Finger + Gaze & 69.5 \%& 7.1° \\
 \hline\hline
 Finger + Gaze + Head & \textbf{73.7 \%} & \textbf{5.2°} \\
 \hline
\end{tabular}
\caption{Modality based results using 12 classes (AOIs 1 - 12)}
\label{Table:monomodal}
\end{table}

\begin{table}[th]
 \begin{tabular}{|c | c c | }
 \hline
 \textbf{Modality} & \textbf{Accuracy} $\uparrow$ & \textbf{MAD} $\downarrow$  \\ 
 \hline\hline
 Head & 38.5 \% & 11.2° \\ 
 \hline
 Gaze & 57.1 \%& 8.2°  \\
 \hline
 Finger & 70.6 \%& 9.7° \\
 \hline\hline
 Gaze + Head & 64.5 \% & 6.8°  \\
 \hline
 Finger + Head & 72.3 \% & 5.9° \\ 
 \hline
 Finger + Gaze & 71.7 \% & 6.6° \\
 \hline\hline
 Finger + Gaze + Head & \textbf{77.6 \%} & \textbf{4.9°} \\
 \hline
\end{tabular}
\caption{Modality based results using 9 classes (AOIs 4 - 12)}
\label{Table:monomodal2}
\end{table}

\begin{table}[th]
 \begin{tabular}{|c | c c | }
 \hline 
 \textbf{Modality} & \textbf{Accuracy} $\uparrow$ & \textbf{MAD} $\downarrow$   \\  
 \hline\hline
 Gaze & 65.0 \% & 7.0°  \\
 \hline\hline
 Finger + Gaze + Head & \textbf{83.9 \%} & \textbf{4.1°} \\
 \hline
\end{tabular}
\caption{Modality based results using 7 classes (AOIs 4-9, 12)}
\label{Table:monomodal3}
\end{table} 

Upon carefully analyzing the videos, we could not find any substantial reason for the poor performance on some users. One of the possible reasons could be very sunny conditions that caused the drivers to clench their eyes slightly. This resulted in poor quality of gaze data, which had some errors in measurements. Another observation was the alternative use of right or left hand for pointing towards the same AOI. This resulted in different pointing angles because of the position of the arm. However, we can not concretely conclude on any of these reasons. A more extensive study needs to be conducted to explain the reasons behind this user behaviour. The differences in recognition of the pointing direction of the different drivers, shows the need for a personalized fusion approach or the implementation of an online adaptive learning approach.

\subsection{Comparison of Different Models}
In this section we compare the different machine learning models. The results are shown in Table \ref{Table:models}. As can be seen, there is no significant difference between the deep neural network models. However, when compared with SVR, there is a big difference of 19\% in accuracy and a difference of 3° in MAD. When analyzing RF regression against CNN, there is only a small difference of 4\% in accuracy and 1° in MAD. From these results, we can see that even with a small dataset, the deep neural networks are able to learn appropriately, and produce sufficiently good results when compared to conventional machine learning.

\begin{table}[t]
 \begin{tabular}{|c || c  c || c |}
 \hline
 \multirow{2}{*}{\textbf{Model}} & \multicolumn{2}{c||}{\textbf{ Linear regression}} & {\textbf{Classification}}\\
 \cline{2-4}
  & \textbf{Accuracy} $\uparrow$ & \textbf{MAD} $\downarrow$  & \textbf{Accuracy} $\uparrow$ \\ 
 \hline\hline
 CNN (base model) & \textbf{73.7 \%} & \textbf{5.1°}  & 73.9 \% \\ 
 \hline
 RNN & 73.3 \% & 5.9°  & 73.0 \%\\
 \hline
 FC-NN & 72.1 \% & 5.5°  & 74.9 \% \\
 \hline
 SVR / SVM* & 55.0 \% & 8.4° & 71.4 \% * \\
 \hline
 RF  & 69.4 \% & 6.1° & \textbf{76.8 \%}   \\ 
 \hline
\end{tabular}
\caption{A comparison of Machine learning models using  linear regression (with the cosine similarity loss for DNNs) and classification (using softmax loss for DNNs)}
\label{Table:models}
\end{table}

\subsection{Processing Speed}
We used an Intel Xeon 16 core processor with a Quadro P5000 GPU for evaluating the models. The processing speeds for all the models are presented in Table \ref{Table:speed}. It is to be noted here that SVR and RF are run on the CPU, while the rest use the GPU. SVR appears to be the fastest, processing about 17,800 frames per second (fps) and taking 0.5 milliseconds (ms) to process a sample of 8 frames. RNN, on the other hand, is the slowest taking 15 ms for one sample. Nonetheless, as the neural networks are not very deep, the processing times are practically applicable and fast, especially when using CNN.


\begin{table}[t]
 \begin{tabular}{|c | c c c|}
 \hline
 \textbf{Model} & \textbf{Time per sample} $\downarrow$ & \textbf{Speed} $\uparrow$ & \textbf{Processor}\\
 \hline\hline
 CNN & 1.1 ms & 7,520 fps & GPU\\ 
 \hline
 RNN & 15.4 ms & 520 fps  & GPU\\
 \hline
 FC-NN & 2.2 ms & 2,800 fps & GPU\\
 \hline
 SVR & 0.5 ms & 17,800 fps  & CPU\\
 \hline
 RF  & 4.2 ms & 1,950 fps  & CPU\\ 
 \hline
\end{tabular}
\caption{Processing time of Machine learning models}
\label{Table:speed}
\end{table}

\subsection{Alternate Approach Using Softmax Loss}

As we are dealing with a classification problem, we also use the softmax loss instead of cosine similarity. Softmax loss function has properties well suited for classification. Similarly, instead of using SVR and RF with the matching algorithm presented in section \ref{matching}, we use Support Vector Machine (SVM) and RF classifier. The results are shown in the last column of Table \ref{Table:models}. There is only a slight difference in the accuracy of all the different models. Angular deviation is no longer applicable as there is no output vector, but rather probabilities for each class. SVM and RF classifier perform better than SVR and RF regression, respectively, because of intrinsic properties suited for classification. In this case, however, we observe RF classifier to outperform CNN in terms of the classification accuracy. This can be associated with the relatively small dataset when using deep neural networks. We also did not tune the hyper-parameters of the neural networks, which may have also have an impact.

While using the softmax function as loss, there is a very slight increase in accuracy, which shows our approach using regression and the matching algorithm is almost equally good. The reason for using regression is scalability and flexibility for future work when we work with Points-of-Interest (POI) on the outside of the car. In such a case, the number of classes are not known beforehand, and such a classification approach would be difficult to realize. 

\section{Conclusion}

In this paper, we presented a unique approach for user interaction inside the car where the driver can select various control modules of the car without a touch based input. To validate our approach, we conducted an experiment in a real but stationary car, and developed a novel approach using different deep neural networks to fuse three modalities for a more robust recognition of the driver's focus of visual attention. Unlike previous research work, we used all the three different modalities, namely finger, head pose and gaze, simultaneously. We have demonstrated that the use of multiple sources of input increases the performance. However, using three modalities instead of two (finger and eye gaze) only results in a negligible enhancement in the recognition accuracy of the driver's selected area-of-interest. 

Furthermore, a comparison of deep learning methods and two conventional machine learning methods (namely SVR and RF) to perform the multimodal fusion is presented. We showed that, even with a considerably small dataset, the deep neural networks are able to learn the weights and produce better predictions for the user's referencing direction than the other two methods.

The work presented in this paper is preliminary and is motivated by the BMW Natural Interaction \cite{bmwnatint}. Based on the results achieved in the experiment, we can conclude that there is much potential with this approach for future user experience applications in the automotive industry. Our future work will focus on extending this to various other use cases, especially in the driving case, with a much larger dataset.

\begin{acks}
This work was done in collaboration with the BMW Group. We are grateful to the BMW colleagues for the assistance in the experimental setup and data collection.
\end{acks}

\bibliographystyle{ACM-Reference-Format}
\balance
\bibliography{main}

\appendix

\end{document}